\documentclass[useAMS,usenatbib]{mn2e}
\usepackage{psfrag,graphicx}
\usepackage{color}
\usepackage{txfonts}
\title[Magnetic field amplification in massive primordial haloes]
  {The small scale dynamo and the amplification of magnetic fields in massive primordial haloes}
\author[Latif et al]
  {M.~A.~Latif,$^1$
  D.~R.~G.~Schleicher,$^1$ 
  W.~Schmidt,$^1$
  J.~Niemeyer$^1$
   \newauthor 
   $^1$ Institut f\"ur Astrophysik, Georg-August-Universit\"at, \\
    Friedrich-Hund-Platz 1, D-37077 G\"ottingen, Germany}
\date{today}

\begin{document}

\bibliographystyle{mn2e}

\label{firstpage}

 \maketitle

\def\na{NewA}%
\def\aj{AJ}%
\def\araa{ARA\&A}%
\def\apj{ApJ}%
\def\apjl{ApJ}%
\def\apjs{ApJS}%
\def\ao{Appl.~Opt.}%
\def\apss{Ap\&SS}%
\def\aap{A\&A}%
\def\aapr{A\&A~Rev.}%
\def\aaps{A\&AS}%
\def\azh{AZh}%
\def\baas{BAAS}%
\def\jrasc{JRASC}%
\def\memras{MmRAS}%
\def\mnras{MNRAS}%
\def\pra{Phys.~Rev.~A}%
\def\prb{Phys.~Rev.~B}%
\def\prc{Phys.~Rev.~C}%
\def\prd{Phys.~Rev.~D}%
\def\pre{Phys.~Rev.~E}%
\def\prl{Phys.~Rev.~Lett.}%
\def\pasp{PASP}%
\def\pasj{PASJ}%
\def\qjras{QJRAS}%
\def\skytel{S\&T}%
\def\solphys{Sol.~Phys.}%

\def\sovast{Soviet~Ast.}%
\def\ssr{Space~Sci.~Rev.}%
\def\zap{ZAp}%
\def\nat{Nature}%
\def\iaucirc{IAU~Circ.}%
\def\aplett{Astrophys.~Lett.}%
\def\apspr{Astrophys.~Space~Phys.~Res.}%
\def\bain{Bull.~Astron.~Inst.~Netherlands}%
\def\fcp{Fund.~Cosmic~Phys.}%
\def\gca{Geochim.~Cosmochim.~Acta}%
\def\grl{Geophys.~Res.~Lett.}%
\def\jcp{J.~Chem.~Phys.}%
\def\jgr{J.~Geophys.~Res.}%
\def\jqsrt{J.~Quant.~Spec.~Radiat.~Transf.}%
\def\memsai{Mem.~Soc.~Astron.~Italiana}%
\def\nphysa{Nucl.~Phys.~A}%
\def\physrep{Phys.~Rep.}%
\def\physscr{Phys.~Scr}%
\def\planss{Planet.~Space~Sci.}%
\def\procspie{Proc.~SPIE}%

%
 \begin{abstract}
{While present standard model of cosmology yields no clear prediction for the initial magnetic field strength, efficient dynamo action may compensate for initially weak seed fields via rapid amplification. In particular, the small-scale dynamo is expected to exponentially amplify any weak magnetic field in the presence of turbulence. We explore whether this scenario is viable using cosmological magneto-hydrodynamics simulations modeling the formation of the first galaxies, which are expected to form in so-called atomic cooling halos with virial temperatures $\rm T_{vir} \geq 10^{4}$ K. As previous calculations have shown that a high Jeans resolution is needed to resolve turbulent structures and dynamo effects, our calculations employ resolutions of up to $128$ cells per Jeans length. The presence of the dynamo can be clearly confirmed for resolutions of at least $64$ cells per Jeans length, while saturation occurs at approximate equipartition with turbulent energy. As a result of the large Reynolds numbers in primordial galaxies, we expect saturation to occur at early stages, implying magnetic field strengths of $\sim0.1$~$\mu$G at densities of $10^4$~cm$^{-3}$.
}

 \end{abstract}


\section{Introduction}
Magnetic fields are ubiquitous in the entire cosmos. They are observed in the local universe  \citep{1996Natur.379...47B,2007A&A...470..539B,2008ApJ...677L..17C,2010ApJ...712..536K} and there is growing evidence for their existence at earlier cosmic times up to $z\sim4$ \citep{2008Natur.454..302B,2008ApJ...676...70K,2009ApJ...706..482M}. Upcoming telescopes like LOFAR\footnote{http://www.lofar.org} and SKA\footnote{http://www.skatelescope.org/} will  enable the detection of magnetic fields at higher redshifts. In present-day galaxies, the field strength ranges from a few to tens of $\rm \mu G$. Their presence plays a vital role in the formation and evolution of many astrophysical systems. It is thus interesting to explore whether magnetic fields may have played a similar role at early cosmic times, particularly during the formation of the first galaxies.

The first magnetic fields may have been created via electroweak \citep{1996PhRvD..53..662B} or quantum chromodynamics phase transitions \citep{1989ApJ...344L..49Q}. Alternatively, they are generated by astrophysical mechanisms like the Biermann battery effect or the Weibel instability during structure formation \citep{1950ZNatA...5...65B,1959PhRvL...2...83W,2003ApJ...599L..57S}. Irrespective of their origin, the strength of magnetic fields predicted by these processes is many orders of magnitude lower than the present-day fields. Hence, they must have been amplified during cosmic history.

The simplest way to increase the magnetic field strength is gravitational compression under the constraint of flux-freezing. For a spherical collapse, it can be shown that $\rm B \propto \rho^{2/3}$  where B is the magnetic field strength and $\rm \rho $ is the gas mass density. A more efficient mechanism is provided by astrophysical dynamos. In particular the small-scale dynamo can amplify the small initial fields on very short timescales \citep{1968JETP...26.1031K,2005PhR...417....1B}. This process converts  turbulent  into magnetic energy, leading to magnetic field strengths approximately in equipartition with the turbulent energy. Numerous theoretical and numerical studies support this idea in the context of cosmological structure formation \citep{2010A&A...522A.115S,2011ApJ...731...62F,2010ApJ...721L.134S,2012ApJ...754...99S,2012ApJ...745..154T,Schoberb}. Modeling this process in numerical simulations is however challenging, as the turbulent eddies need to be well resolved with at least $32$ cells to provide a sufficiently large magnetic Reynolds number Rm \citep{2005PhR...417....1B,2010ApJ...721L.134S,2011ApJ...731...62F}. The magnetic Reynolds number is given as Rm$=VL/\mu$, with $V$ and $L$ the characteristic turbulent velocity and length scale, and $\mu$ the magnetic diffusivity. The critical magnetic Reynolds number for dynamo action ranges from $\sim100$ for Kolmogorov to more than $3000$ for Burgers-type turbulence \citep{2012ApJ...754...99S}.

According to the current paradigm of structure formation, the first stars, so-called Pop III stars, were formed in mini-haloes with masses above $\rm \geq 10^{5}~M_{\odot}$ at redshift 30 \citep{2002Sci...295...93A,2004ARA&A..42...79B, 2008Sci...321..669Y,2011Sci...331.1040C}. For these haloes, efficient magnetic field amplification was demonstrated in various studies \citep{2010ApJ...721L.134S,2011ApJ...731...62F,2012ApJ...745..154T,2012ApJ...754...99S}. More massive primordial haloes with virial temperature $\rm \geq 10^{4}$ K were assembled around $\rm z=15-20$ and have typical masses of $\rm \geq 10^{7}~M_{\odot}$ \citep{2008MNRAS.387.1021G,2009Natur.459...49B}. The main coolant in these massive haloes is the Lyman alpha radiation so they are known as atomic cooling haloes. The study of these haloes is of extreme interest as they are presumed to host the first protogalaxies which were the main drivers for cosmic reionization as well as predecessors of present day galaxies. Exploring the magnetic field amplification in these haloes is more challenging, as a larger range of scales need to be bridged, and because of the higher Mach numbers involved in the problem.

These atomic cooling haloes are considered as the potential birthplaces for the formation of supermassive black holes at $\rm z\geq 6$. The latter are presumed to power quasars and have masses of $\rm \geq 10^{9}~M_{\odot}$ \citep{2003AJ....125.1649F,2006AJ....131.1203F,2011Natur.474..616M}. It is not yet fully comprehended how such objects could form in such a short span of time. Different mechanisms have been suggested for the formation of supermassive black holes \citep{1978Obs....98..210R,2004Natur.428..724P,2007MNRAS.374.1557J,2008arXiv0803.2862D,2009MNRAS.tmp..640R,2009ApJ...702L...5B,2009MNRAS.396..343R,2010A&ARv..18..279V,2012MNRAS.421.1465D,2012arXiv1203.6075H}. Given the difficulties with other mechanisms, direct collapse has emerged as the most plausible scenario leading to black hole masses of $\rm 10^{4}-10^{5}~M_{\odot}$ \citep{2002ApJ...569..558O,2003ApJ...596...34B,2006ApJ...652..902S,2006MNRAS.370..289B,2008MNRAS.391.1961D,2008arXiv0803.2862D,2010MNRAS.402.1249S,2010ApJ...712L..69S,Latifa,Latifd,2012MNRAS.425.2854A,2012arXiv1211.0548J,2010A&ARv..18..279V,2012RPPh...75l4901V,2012arXiv1203.6075H}.

The essential condition for the direct collapse scenario is to avoid fragmentation. Molecular hydrogen is the only coolant in zero metallicity haloes which can bring the temperature of the gas down to a few hundred Kelvin and may induce fragmentation by reducing the Jeans mass \citep{2003ApJ...592..975L}. The existence of a strong UV Lyman-Werner flux photo-dissociates $\rm H_{2}$ molecules, suppresses cooling and consequently may halt the fragmentation of a gas cloud. We consider a background  UV flux with a thermal spectrum of $\rm T_{rad}=10^{5}$ K which quenches $\rm H_{2}$ formation via the Solomon process. These conditions create a suitable environment for the growth of  black holes. The presence of magnetic fields may favor this scenario by transferring angular momentum and exerting magnetic pressure depending on the ambient field strength. Hence, exploring the magnetization of atomic cooling haloes is intimately associated with our understanding of the first galaxies and black hole formation. Indeed, strong rotation 
measures have been observed in quasars out to at least redshift 5, providing a further indication that magnetic fields are relevant early on \citep{2012arXiv1209.1438H}, while detailed observations of the field structure in local AGN imply significant contributions to the transport of angular momentum \citep{Beck05}.

In this study, we aim to focus on the magnetization of massive primordial haloes, particularly the amplification of magnetic fields by the small scale dynamo. To accomplish this goal, we perform high resolution cosmological magnetohydrodynamical simulations with resolutions up to 128 cells per Jeans length. We also explore the saturation of dynamo generated magnetic fields due to the small scale dynamo for a Jeans resolution of 64 cells and examine its impact on the morphology of the halo. This study will enable us to comprehend the amplification of magnetic fields by the small scale dynamo in atomic cooling haloes and their potential role in the formation and growth of supermassive black holes.

This article is organized in the following way. In the second section, we discuss the details of numerical techniques used in this study. In the third section, we discuss the results obtained.  In section four, we explain the implications of magnetic fields in black hole formation. In the last section, we summarize the main findings and confer our conclusions.

\section{Numerical Techniques and simulation setup}

The results presented in this work are obtained using the public version of the Enzo code. Enzo is an adaptive mesh refinement (AMR), parallel, grid-based cosmological magnetohydrodynamics code \citep{2004astro.ph..3044O,2007arXiv0705.1556N}. The magnetohydrodynamics (MHD) equations are solved using the Harten-Lax-van Leer (HLL) Riemann solver with piece-wise linear construction. We enforce $\rm \nabla \cdot B =0$ constraint employing the Dedner scheme \citep{2008ApJS..176..467W,2010NewA...15..581W}. 

The simulations are commenced with cosmological initial conditions generated from Gaussian random fields. We employ the inits package available with the public version of the Enzo code to create nested grid initial conditions. Our simulations start at redshift $\rm z=99$ with a top grid resolution of $\rm 128^{3}$ cells and we select the massive halo of $\rm 8 \times 10^{6}~M_{\odot}$ at redshift 15 using the halo finder of \cite{2011ApJS..192....9T}. Two initial nested levels of refinement are subsequently added each with a resolution of $\rm 128^{3}$ cells. Our simulation box has a cosmological size of 1 Mpc $\rm h^{-1}$ and is centered on the massive halo. In total, we initialize 6291456 particles to compute the evolution of the dark matter dynamics and have a final dark matter resolution of 300 $\rm M_{\odot}$. The parameters for creating the initial conditions and the distribution of baryonic and dark matter components are taken from the WMAP seven years data \citep{2011ApJS..192...14J}. We 
further allow additional 27 levels of refinement in the central 62 kpc region of the halo during the course of simulation. It gives us a total effective resolution of 0.17 AU at $\rm z=11.5$. The resolution criteria used in these simulations are based on the Jeans length, the gas overdensity and the particle mass resolution. The grid cells matching any of these requirements are marked for a refinement.  

The simulations conducted in this work mandated Jeans length resolutions of 16, 32, 64 and 128 (hereafter called $\rm J_{16}$, $\rm J_{32}$, $\rm J_{64}$, $\rm J_{128}$) cells throughout their evolution. 
To include the primordial non-equilibrium chemistry, the rate equations of the following 9 species: $\rm H,~H^{+},~He,~He^{+},~He^{++},~e^{-},~H^{-},~H_{2},~H_{2}^{+}$ are self-consistently solved in the cosmological simulations. We make use of the $\rm H_{2}$ photo-dissociating background UV flux implemented in the Enzo code. An external UV field of constant strength $\rm 10^{3}$ in units of $\rm J_{21}$ is used in the simulations. We presume that such flux is generated from a nearby star forming halo \citep{2008MNRAS.391.1961D} and is emitted by Pop III stars with a thermal spectrum of $\rm 10^{5}$ K. We include several cooling and heating mechanisms like collisional ionization cooling, radiative recombination cooling, collisional excitation cooling, H$_{2}$ cooling as well as $\rm H_{2}$ formation heating. The chemistry solver used in this work is a modified version of \cite{1997NewA....2..181A,1997NewA....2..209A}. We stop our simulations after they reach the maximum refinement level. The results at later 
stages would not be reliable without providing additional means to ensure the Truelove criterion.


\begin{figure*}
\hspace{-4.0cm}
\centering
\begin{tabular}{c}
\begin{minipage}{12cm}
\includegraphics[scale=0.8]{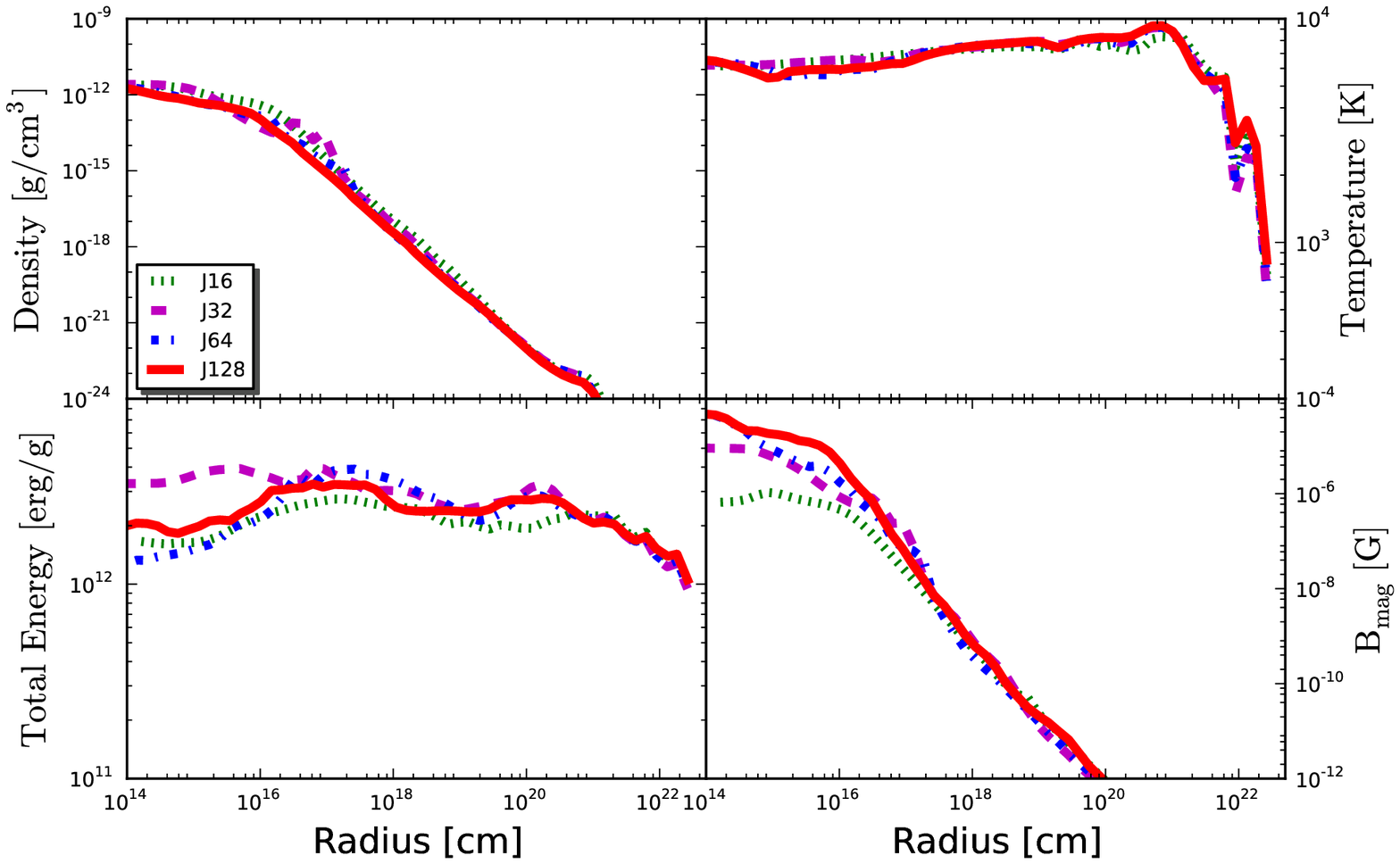}
\end{minipage}
\end{tabular}
\caption{This figure shows the radially binned spherically averaged radial profiles for the various Jeans resolutions as shown in the legend. The upper left panel shows the density radial profiles. The temperature radial profiles are depicted in the upper right panel. The bottom left panel shows the total energy radial profiles. The radial profiles for the total strength of magnetic fields are shown in the bottom right panel.}
\label{fig}
\end{figure*}

\begin{figure*}
\hspace{-2.0cm}
\centering
\begin{tabular}{c}
\begin{minipage}{12cm}
\includegraphics[scale=0.6]{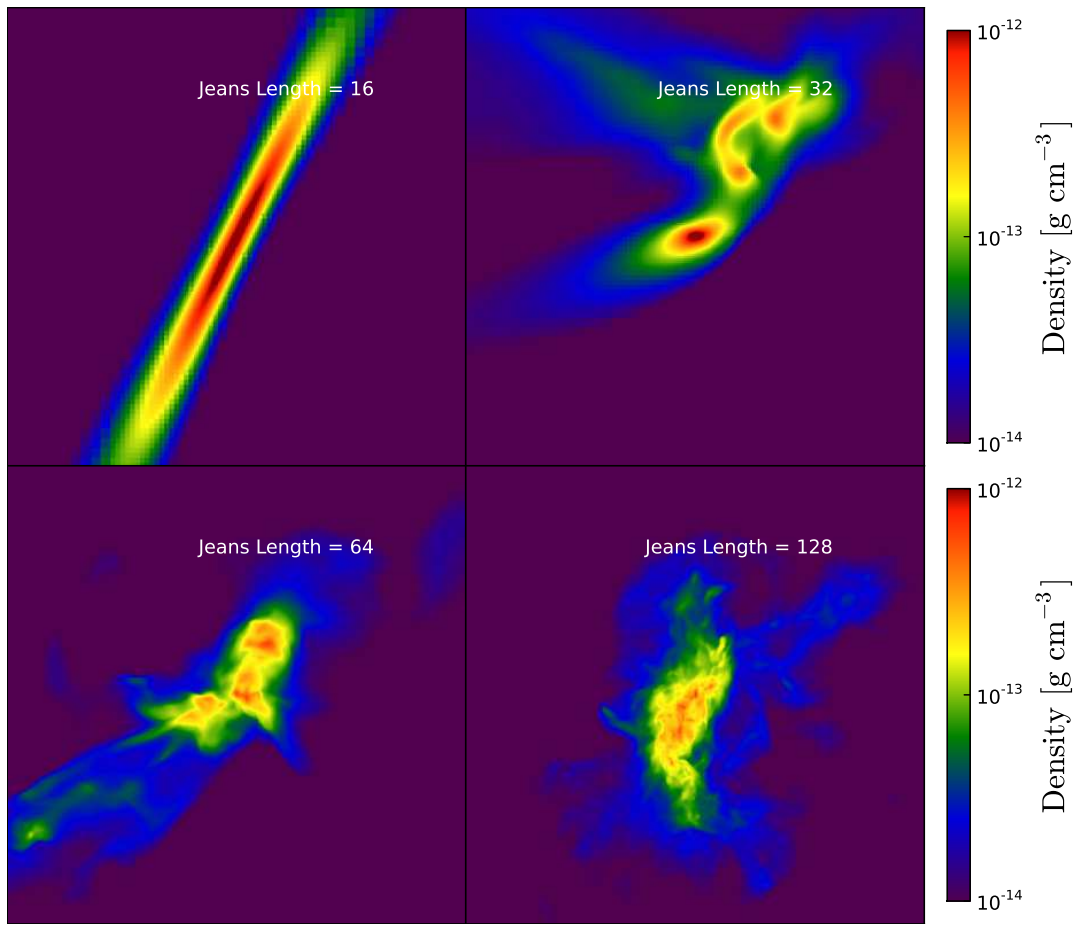}
\end{minipage}
\end{tabular}
\caption{The figure illustrates the state of simulations at the collapse redshift. Density projections are shown for the central 4000 AU region of the halo for various Jeans resolutions. The top panels show the cases for 16 (left) and 32 (right) cells per Jeans length. The bottom panels depict the cases for Jeans resolutions of 64 and 128 cells from left to right respectively.}
\label{fig0}
\end{figure*}

\begin{figure*}
\hspace{-4.0cm}
\centering
\begin{tabular}{c}
\begin{minipage}{12cm}
\includegraphics[scale=0.6]{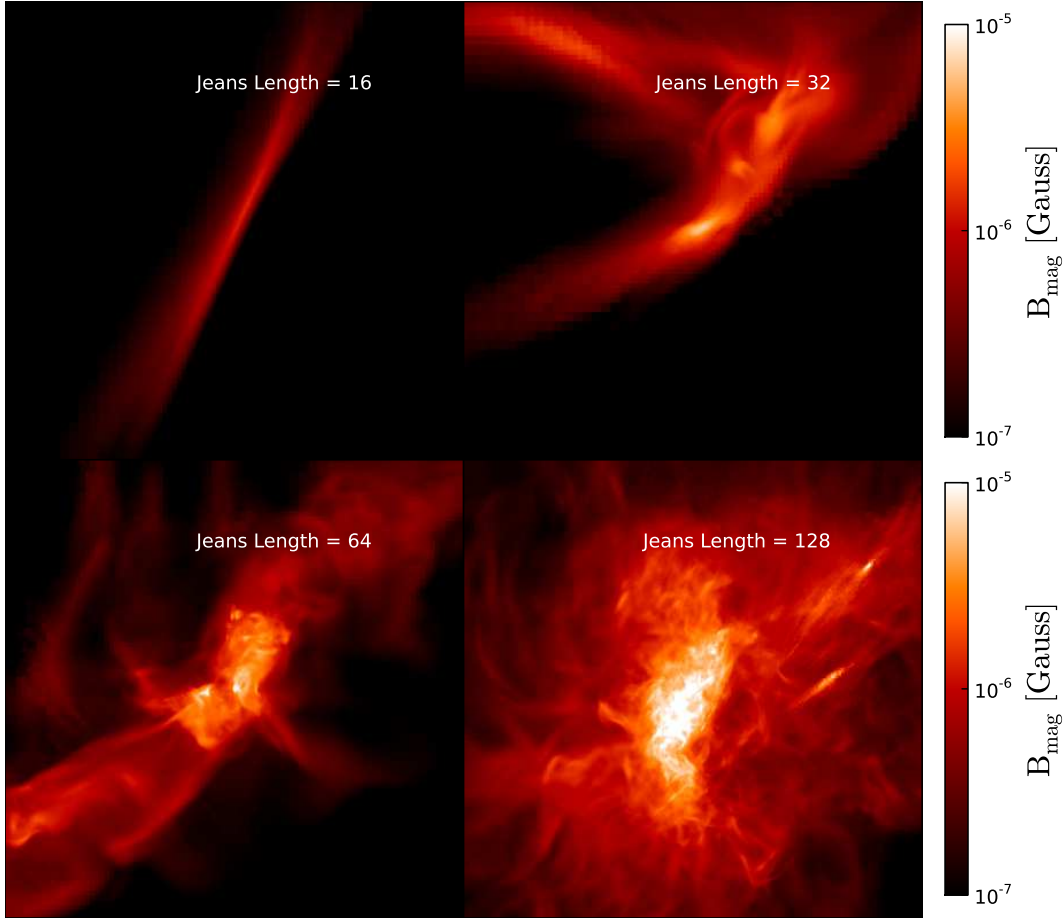}
\end{minipage}
\end{tabular}
\caption{ Density-weighted magnetic field strength projections for various Jeans resolutions are shown in this figure. They are depicted at the central field of view of 4000 AU. Top panels show the cases for 16 (left) and 32 (right) cells per Jeans length. Bottom panels depict the cases for Jeans resolutions of 64 and 128 cells from left to right respectively. }
\label{fig1}
\end{figure*}
\begin{figure*}
\hspace{-1.0cm}
\centering
\begin{tabular}{c c}
\begin{minipage}{6cm}
\includegraphics[scale=0.45]{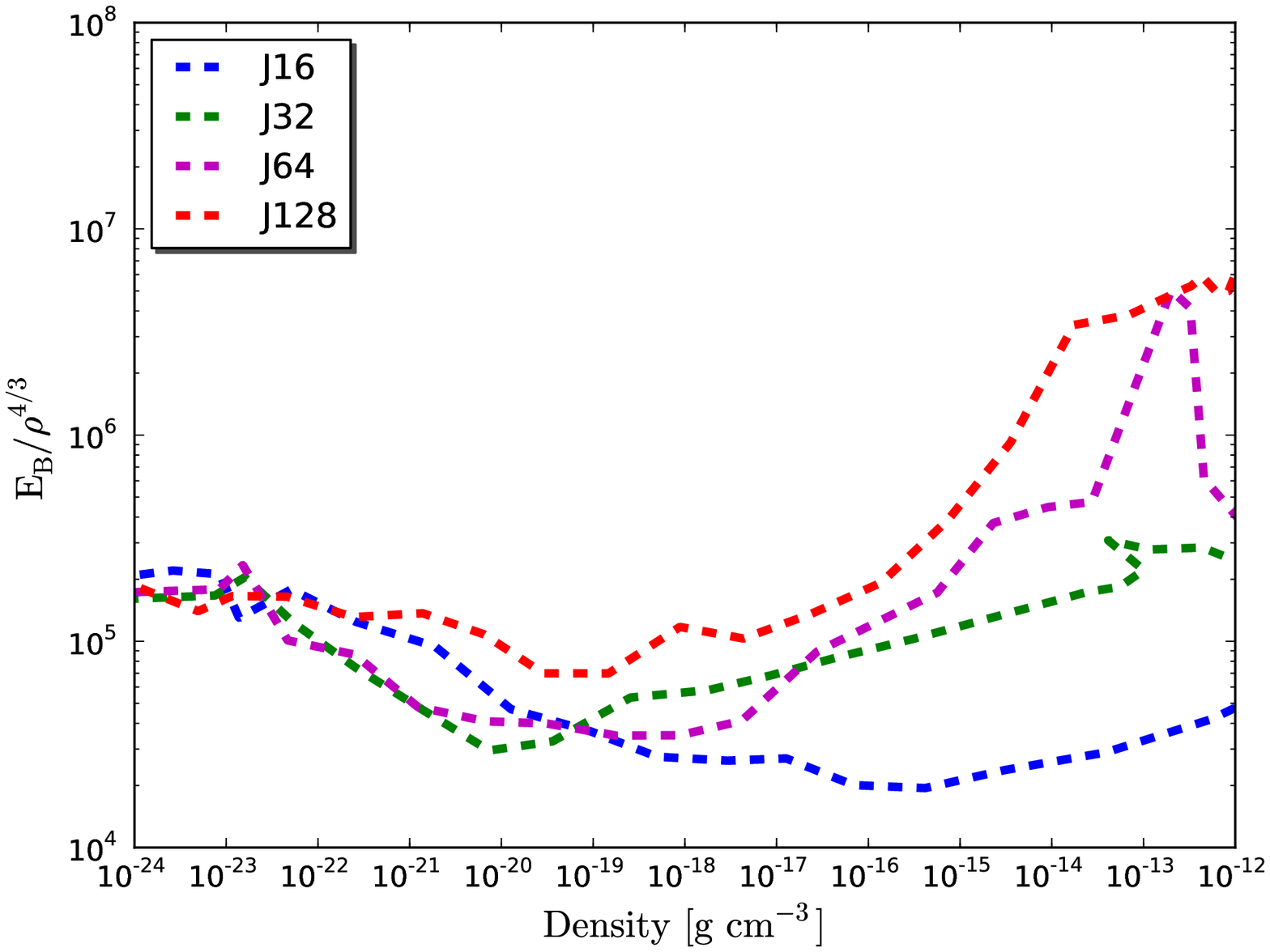}
\end{minipage} &
\hspace{3cm}
\begin{minipage}{8cm}
\includegraphics[scale=0.45]{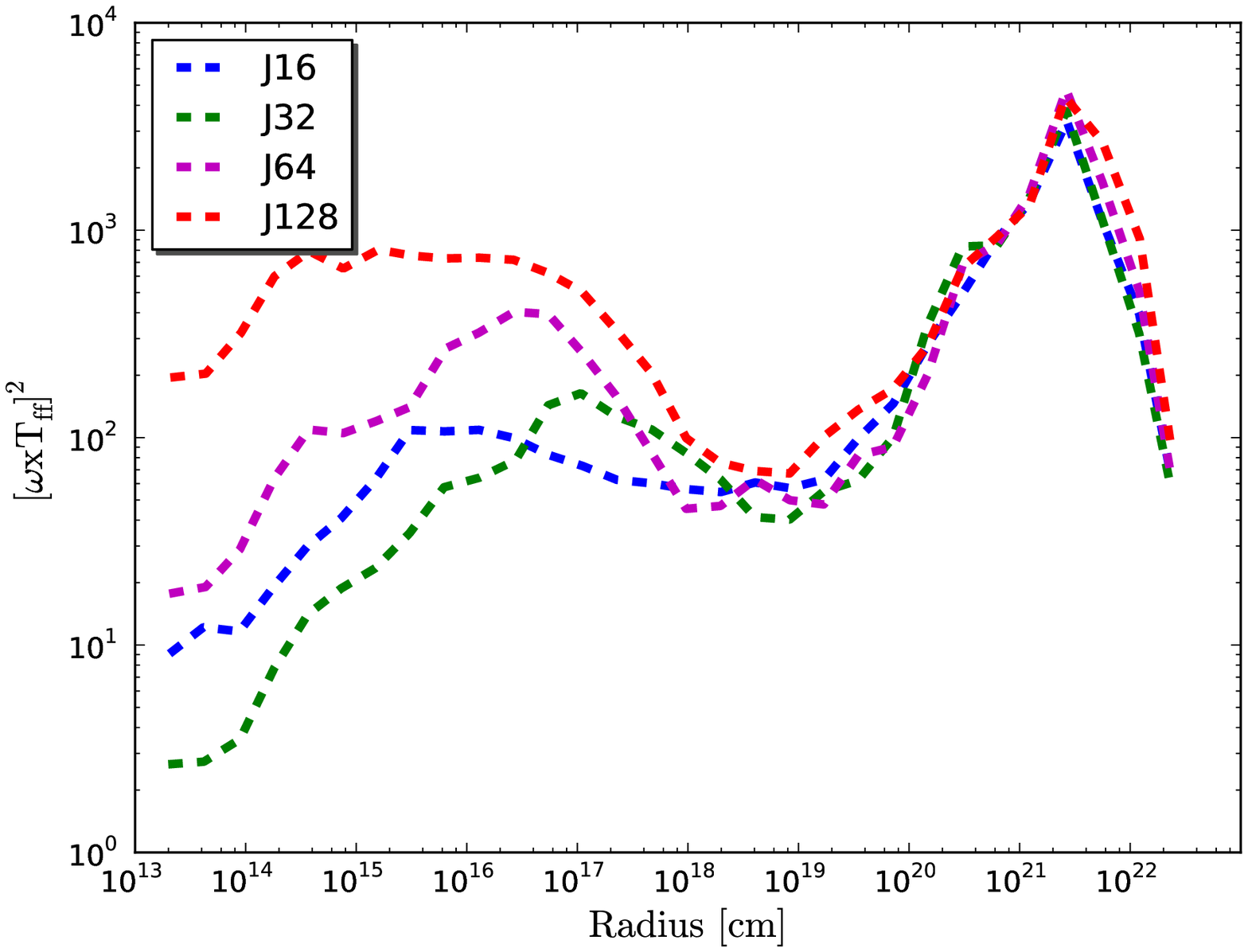}
\end{minipage}
\end{tabular}
\caption{ In the left panel of the figure, magnetic energy scaled by $\rm \rho^{4/3}$ is plotted against the gas density for various Jeans resolutions as indicated in the legend.  Radial profiles of the vorticity normalized by the free-fall time are shown in the right panel.}
\label{fig2}
\end{figure*}

 \begin{figure}
\centering
\includegraphics[scale=0.4]{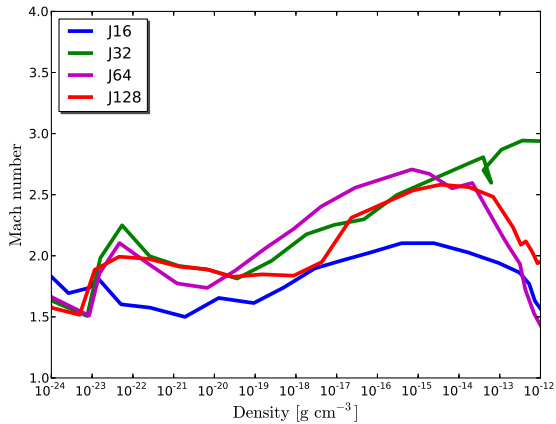}
\caption{In this figure Mach number is plottd against the gas density. Different colors represent various resolutions as indicated in legends}
\label{fig30}
\end{figure}

\begin{figure*}
\hspace{-4.0cm}
\centering
\begin{tabular}{c c}
\begin{minipage}{6cm}
\includegraphics[scale=0.45]{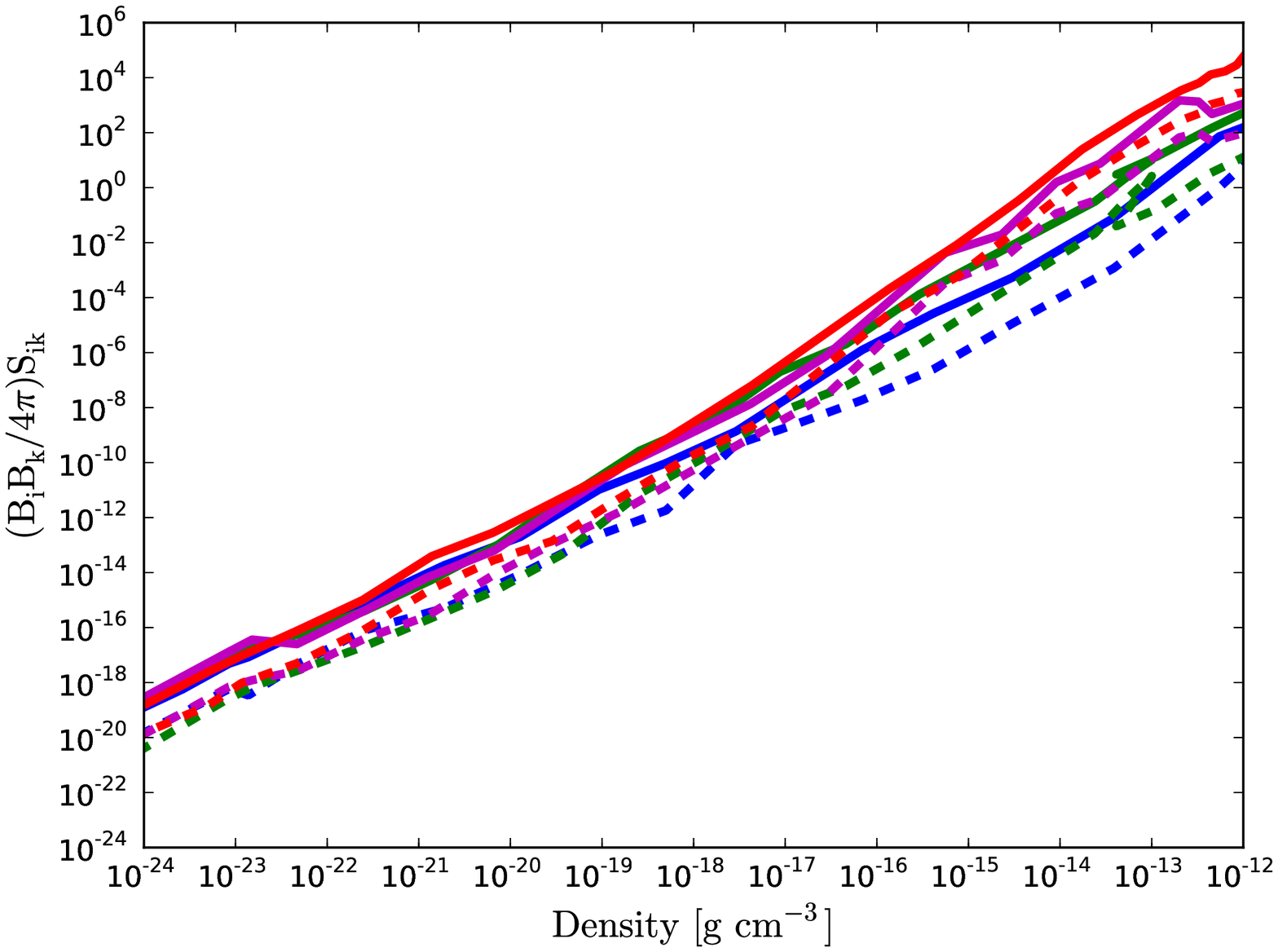}
\end{minipage} &
\hspace{2cm}
\begin{minipage}{6cm}
\includegraphics[scale=0.45]{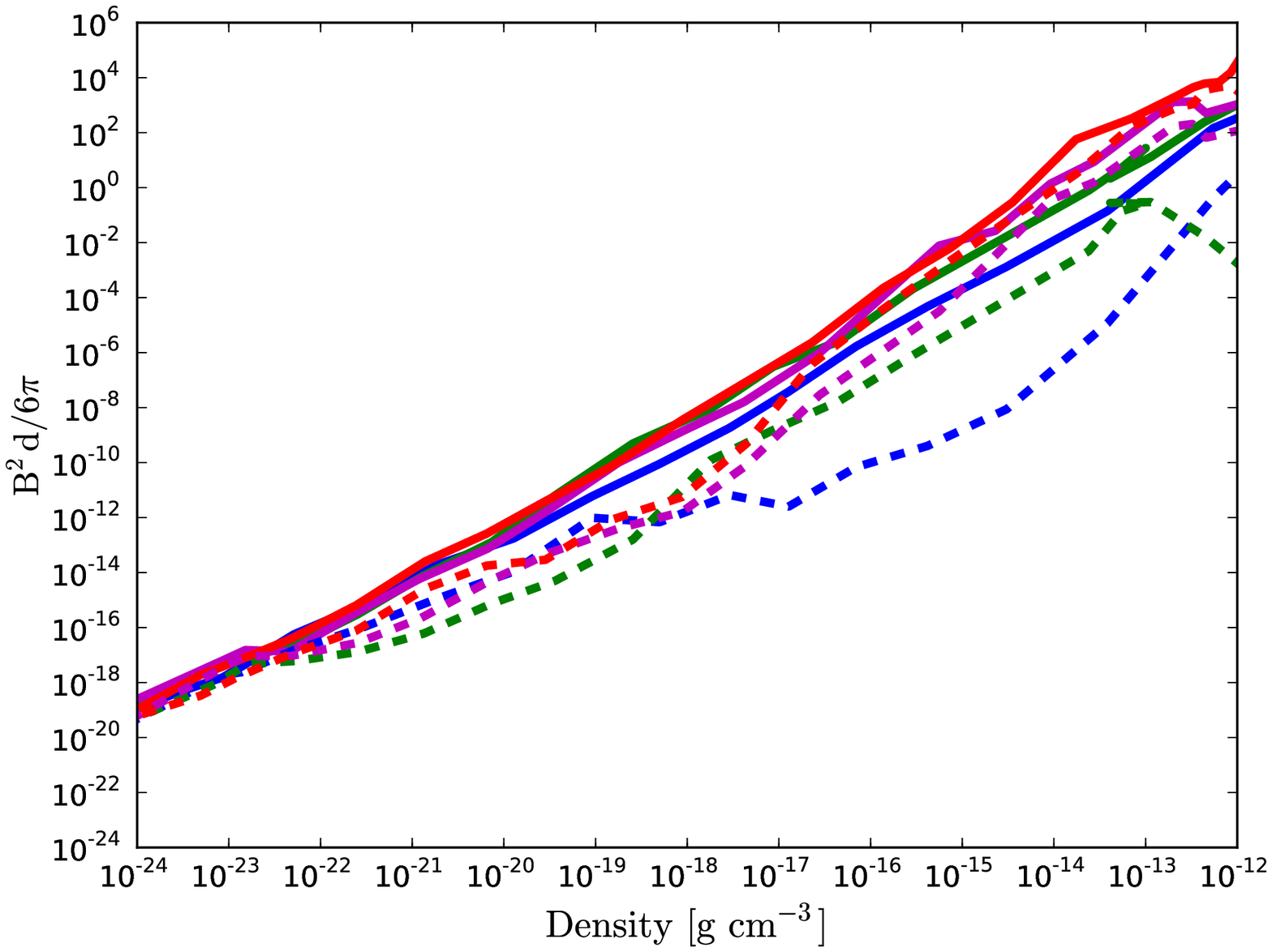}
\end{minipage}
\end{tabular}
\caption{This figure depicts the amplification of magnetic field strength by shear (left) and compression (right). The solid lines represent the contribution of positive components while dotted lines stand for the negative components of both shear and compression. Blue, green, purple and red colors represent the Jeans resolutions of 16, 32, 64 and 128 cells respectively.}
\label{fig3}
\end{figure*}

\begin{figure*}
\hspace{-4.0cm}
\centering
\begin{tabular}{c}
\begin{minipage}{6cm}
\includegraphics[scale=0.45]{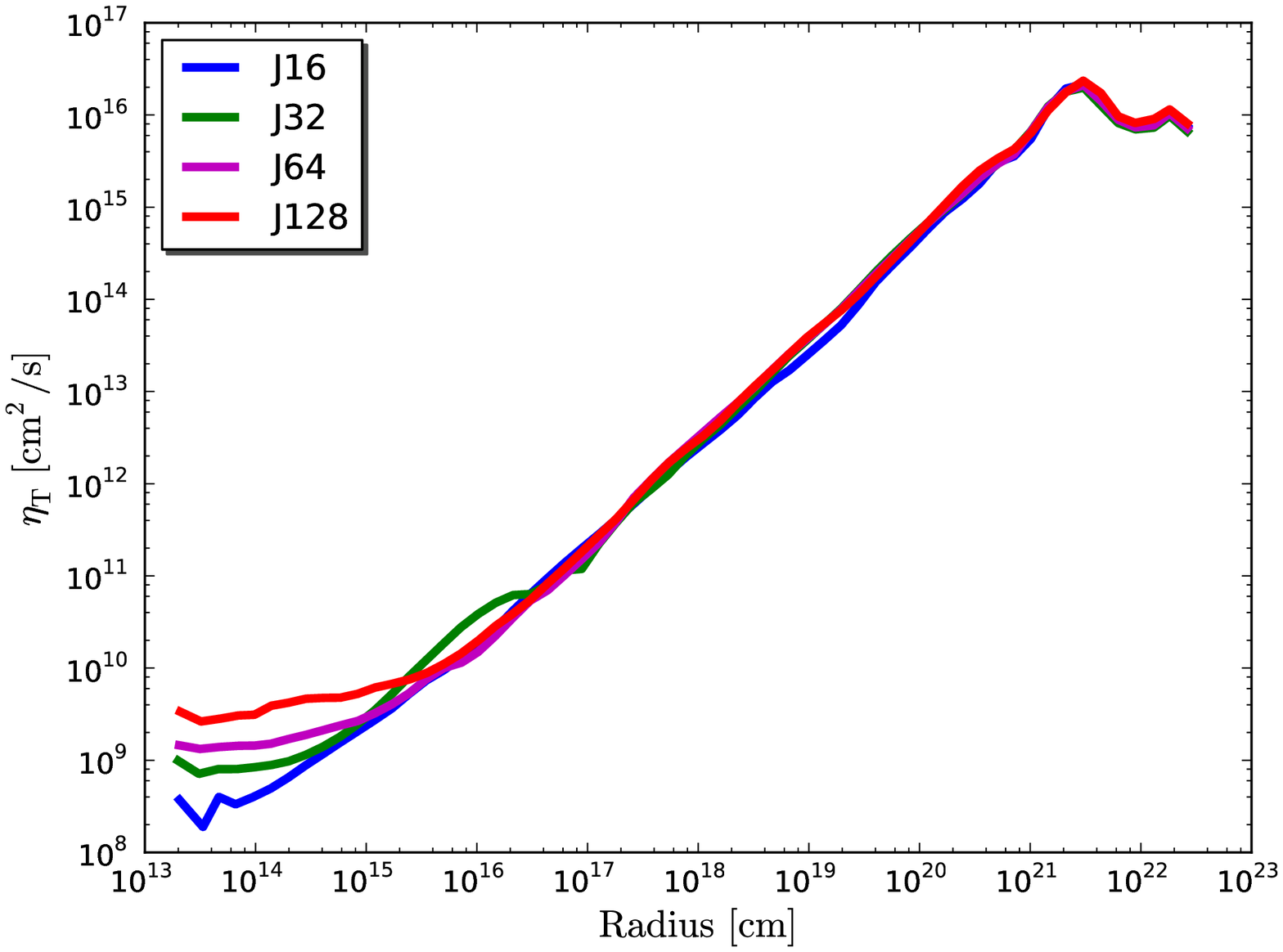}
\end{minipage}
 \end{tabular}
\caption{The radial profile of turbulent viscosity is shown in the figure. Different colors represent various resolutions as indicated in legends.}
\label{fig4}
\end{figure*}

\section{Results}
In this section, we present the results obtained from cosmological MHD simulations for a halo of $\rm 3 \times 10^{7}~M_{\odot}$ at redshift 11.5 selected from the sample of \cite{2012arXiv1210.1802L}. The density perturbations begin to collapse under gravitational instability at a redshift around 30. They merge with each other to form a halo in accordance with the hierarchical paradigm of structure formation. Gas falls into the dark matter potential and gets shock heated in the nonlinear phase of the collapse. Consequently, the temperature of the halo starts to increase. During the process of virialization, part of the gravitational energy is converted into thermal and kinetic random motions. This process continues until the halo reaches the state of virial equilibrium.

\subsection{Resolution study}

\subsubsection {Radial profiles}

The radial profiles of the halo for different resolutions per Jeans length and the same central density are shown in figure \ref{fig}. The maximum density reached in the simulation is about $\rm 10^{-12}~g~cm^{-3}$. The averaged density radial profile is shown in the top left panel of figure \ref{fig}. It can be seen that the density follows an $\rm R^{-2}$ behavior which is expected for close to isothermal collapse. The small bumps in the profiles for the $\rm J_{32}$ and $\rm J_{64}$ cases indicate the presence of clumps at a distance of 0.1 pc. The flat central region in the profile corresponds to the Jeans length below which the effect of gravity is negligible. The overall density profiles agree on the scale above 0.1 pc although there are differences in the interior, which are likely related to the morphological differences discussed below.

The upper right panel of figure \ref{fig} depicts the temperature radial profile. The temperature initially rises at larger radii due to the virialization shocks and reaches a maximum around $\rm 10^{4}$ K which corresponds to the virial temperature of the halo. The cooling due to Lyman alpha radiation becomes important at these temperatures and brings the temperature of the halo down to about 7000 K. In the presence of strong UV Lyman Werner flux $\rm J_{21}=10^{3}$ the formation of $\rm H_{2}$ remains inhibited and the halo collapses almost isothermally. It is found that the thermal properties of the halo are independent of the Jeans resolution. The total energy radial profile is shown in the bottom left panel of figure \ref{fig}. The value of total energy is a few times $\rm 10^{12}~erg/g$ which is comparable to our previous study. There is an increase in the total energy at outer radii due to the infall of the gas. It becomes almost constant at intermediate radii. The variations in the core may arise from the differences in the morphology of the halo. On scales above 0.1 pc, the total energy is converged for different resolutions.

The averaged radial profile of the magnetic field strength is presented in the bottom right panel of the figure \ref{fig}. The simulations were seeded with the same initial magnetic field strength of $\rm 10^{-14}~G$ at the initial redshift. The values of the magnetic field follow a power law behavior for different resolutions at larger radii. The magnetic field amplification seems to proceed via gravitational compression at larger radii while turbulent amplification kicks in in the interior. The maximum value of the magnetic field strength is about $\rm 10^{-4}$ G. The value of $\rm B_{mag}$ is about two orders of magnitude higher for the $\rm J_{128}$ case compared to the $\rm J_{16}$. This increase in magnetic field strength is a consequence of the turbulent amplification of the magnetic field as turbulent eddies are better resolved in the highest resolution case. We will focus on the details of amplification in the next subsection.

\subsubsection{Morphology of the halo}

The state of the simulations at the collapse redshift for different Jeans resolutions is illustrated in the density projections of figure \ref{fig0}. Remarkable dissimilarities in the morphology of the halo are found for different resolutions. For the $\rm J_{16}$ case, the halo is collapsed into a single filament. There are no turbulent structures present in this case which indicates that turbulence is clearly under-resolved. The more complex clumps are observed for a Jeans resolution of $\rm \geq$ 32 cells. They are very well separated from each other and may form binary or even multiple systems. Again, turbulence seems to be marginally resolved in this case. For the Jeans resolution of 64 cells, the medium inside the halo becomes turbulent. One can see that again there are multiple clumps and turbulence is better resolved compared to the previous cases. They are not gravitationally bound and have sub-solar masses.

For the highest resolution case, turbulence seems to be very well resolved. There are no coherent structures present in the halo. Our results suggest that one needs a Jeans resolution of at least $\rm \geq 64$ cells to resolve gravity driven turbulence in atomic cooling haloes. In the highest resolution case, the halo has become highly turbulent. This can have important implications for the ultimate fate of the halo. Turbulence can efficiently transport angular momentum and may lead to the formation of a seed black hole or a supermassive star \citep{2009ApJ...702L...5B}. Our results show structure of the halo strongly varies with resolution and the presence of turbulence may lead to different fragmentation properties.

\subsubsection{Magnetic field Amplification and turbulence}

The simulations were started with the same initial seed magnetic field of $\rm 10^{-14}$ G (proper) at redshift 99. The magnetic field is amplified during the collapse, reaching a peak value of 100 G at densities of $\rm 10^{-12}~g/cm^{3}$ at $\rm z=11.5$. In total, the magnetic field is amplified about 10 orders of magnitude. This is evident in figure \ref{fig1} where the total strength of the magnetic field is depicted for different Jeans resolutions. The strength of magnetic field is correlated with the density structure present inside the halo. The amplification of magnetic fields for $\rm J_{16}$ and $\rm J_{32}$ cases is mainly due to the gravitational compression as the turbulence is poorly resolved. 

In fact, we clearly recognize turbulent structures for Jeans resolutions of $\rm \geq 64$ cells. The turbulent energy is converted into magnetic energy via the small scale dynamo which exponentially amplifies the magnetic field in addition to the contribution from flux freezing and gravitational compression. The large region with higher values of vorticity in the $\rm J_{128}$ case enhances the magnetic energy which results in enhanced magnetic field strengths as shown in the bottom right panel of figure \ref{fig1}. The magnetic field strength is higher in the center as turbulence is produced locally due to the gravitational collapse. Our results clearly demonstrate the necessity of a Jeans length resolution of $\geq$ 64 cells to excite the dynamo action. The requirement of a higher Jeans resolution compared to \citet{2011ApJ...731...62F} may arise due to the differences in the numerical schemes. In particular, we note that the HLLR solver is somewhat more diffusive compared to the HLL3R employed by \citet{2011ApJ...731...62F}.

We have further quantified the contribution of the dynamo generated magnetic field as manifested in the left panel of figure \ref{fig2}. The magnetic energy would scale as $E_{B}=\rho^{4/3}$ according to the expectation of spherical collapse. We infer the dynamo amplification of the magnetic field by dividing the magnetic energy by the maximum possible contribution from the frozen magnetic field lines. It can be noticed from the plot that there is no contribution of dynamo action for a $\rm J_{16}$ run and magnetic energy seems to be dissipated. For higher resolution runs, initially there is a slight decrease in magnetic energy, which can be due to the strong virialization shocks that destroy the coherence in turbulence and magnetic fields necessary to derive the dynamo (as interpreted by \cite{2011ApJ...731...62F,2012MNRAS.423.3148S}). Later on the contribution of the small scale dynamo becomes important and efficiently amplifies the magnetic field. For the best resolution case ($\rm J_{128}$), dynamo amplification is about two orders of magnitude above the flux freezing.

In figure \ref{fig30} the Mach number is plotted against the gas density for different Jeans resolutions. The Mach number peaks around 3. For the highest resolution case, the maximum Mach number in the core of the halo is 2. Overall, the turbulence inside the halo is supersonic. The small scale dynamo is known to be less efficient for compressible turbulence \citep{2011PhRvL.107k4504F,2012ApJ...754...99S}, as observed in the halos modeled here. The amplification of the magnetic field is therefore less efficient compared to the minihaloes explored in previous studies  \citep{2010ApJ...721L.134S,2011ApJ...731...62F,2012ApJ...745..154T,2012ApJ...754...99S}. However, the latter is likely compensated in astrophysical systems by the much higher Reynolds numbers, as the growth rate scales as $\rm Re^{1/3}-Re^{1/2}$ \citep{2012ApJ...754...99S}.

Additional evidence for the turbulent amplification of magnetic fields is shown in the right panel of figure \ref{fig2}. The radial profile of vorticity normalized by the local free-fall time demonstrates that the amount of vorticity increases with resolution particularly in the center of the halo. The relatively high vorticity in the inner region indicates turbulence driven by the collapse. The peak in vorticity at larger radii is because of virialization shocks.
  
Gravitational compression squeezes magnetic field lines under the condition of flux freezing for the ideal MHD conditions. The rate of change of magnetic pressure can be computed from the induction equation \cite{2013arXiv1302.4292S}
 
\begin{equation}
\rm {D \over Dt} \left( {B^{2} \over 8 \pi} \right) =  {1 \over 4 \pi}\left(B_{i}B_{j}S_{ij}^{*} - 2/3B^{2}d \right),
\label{Bpres}
\end{equation}
where $\rm{D \over Dt}$ is $\rm {\partial \over \partial t} + v.\nabla$, $d=\nabla \cdot v$ is the divergence of the velocity and $\rm S_{ij}^{*}=S_{ij}-1/3\delta_{ij}$ is the trace-free rate of strain tensor. The first term on the right hand side of equation \ref{Bpres} comes from the action of turbulent shear on magnetic fields while the second term results from the gas compression both due to gravity and shocks. In each case the positive contribution exceeds the negative contribution, implying a net amplification effect. We have computed the contributions of both shear and compression for different resolutions and they are plotted in figure \ref{fig3}. We see a clear trend of enhanced amplification rate with increasing resolution. The amplification by local shear and compression is quite comparable. This has already been seen in MHD turbulence simulations and seems to be a general property of self-gravitating turbulence. Moreover, amplification correlates very well with density as expected in gravity driven turbulence of collapsing halos \cite{2013arXiv1302.4292S}.

As shown by \citet{Roberts75}, the amplification of magnetic fields via the small-scale dynamo on a fixed spatial scale can also be understood in terms of the mean electromotive force acting on that scale, given as
\begin{equation}
\langle \vec{v}\times\vec{b}\rangle=-\frac{1}{2}\nabla\nu_T\times\vec{B}-\nu_T\nabla\times\vec{B},
\end{equation}
with $\nu_T$ the turbulent viscosity. A complementary understanding of the dynamo may thus be obtained from the evolution of the turbulent viscosity during the simulation. Here, we pursue only an order of magnitude estimate as
\begin{equation}
\rm \nu_{turb}= v_{turb}\lambda_{J}
\label{Tvis}
\end{equation}
with $\rm \lambda_{J}$  the Jeans length and $\rm v_{turb}$ the turbulent velocity. The Jeans length is thus adopted as the turbulent length scale, as local collapse is driving the turbulent motions. To estimate  the turbulent velocity, we separate it from the radial and rotational velocity via
\begin{equation}
\rm v_{turb} = \sqrt{v_{rms}^{2} -v_{rad}^{2} -v_{rot}^{2}},
\label{vturb}
\end{equation}
where $\rm v_{rms}$ is the root mean square velocity, $\rm v_{rad}$ is the radial velocity and $\rm v_{rot}$ is the rotational velocity.
The radial profile of the turbulent viscosity is depicted in figure \ref{fig4}, showing a power law behavior regardless of the resolution which decreases towards the smaller radii. Although one would expect the turbulent viscosity to decrease with increasing resolution, the behavior is more complicated. The amount of viscosity is enhanced in the center for higher resolution runs because of better resolved turbulent structures. The maximum value of the turbulent viscosity is few times $\rm 10^{9}~cm^{2}/s$ for the best resolution case. The presence of such turbulent viscosity thus regulates the growth of the dynamo. It has also important implications for the angular momentum transport in such halos, either in potential accretion disks or more extended turbulent structures. While a detailed discussion is beyond the scope of this article, it provides an order of magnitude estimate for the turbulent viscosity which can be employed for follow-up calculations.

.

\subsection{Saturation of the magnetic field}

Only a modest range of Reynolds numbers can be achieved in numerical simulations due to the exorbitant computational costs. Simulations starting from weak fields thus will not achieve saturation, as they will still underestimate the growth rate (see \cite{Schoberb}). Saturation is expected to occur when backreactions from non-ideal MHD effects or the Lorentz force become important. The central question concerning the amplification of dynamo generated magnetic fields is at what field strength saturation takes place. In numerical simulations of dynamo generated fields, saturation occurs when the magnetic energy reaches a fraction of the equipartition value \citep{2005PhR...417....1B,2011PhRvL.107k4504F}. The saturated field strength depends on the turbulent energy injection scale, the type of driving and the Mach number of the turbulence.

To explore the appropriate saturation level via numerical simulations is rather challenging, as numerical resolution limits the growth rate of magnetic fields, and a clear assessment of saturation is only possible through the comparison of unsaturated and saturated runs. We therefore vary the initial field strength to check for which magnetic seed field we observe a different behavior of $E_{mag}/E_{kin}$ (implying saturation). While this is the only feasible approach in such a cosmological scenario, we note that the interpretation is not always straightforward, as potential uncertainties need to be taken into account. For instance, when varying the initial field strength, even small backreactions from the Lorentz force may give rise to a different realization of (statistically) the same turbulent velocity field. This implies that the morphology appears somewhat different. The resulting fluctuations imply that the evolution of the magnetic field is never identical and needs to be accounted for. Even further, it is conceivable that the presence of a moderate field induces changes in the dynamics, potentially affecting the amount of turbulence and thus the appropriate saturation level. For the procedure described here, it is thus important to note that only with consistent results from a series of runs, a solid conclusion can be pursued. 

\begin{figure*}
\centering
\includegraphics[scale=0.8]{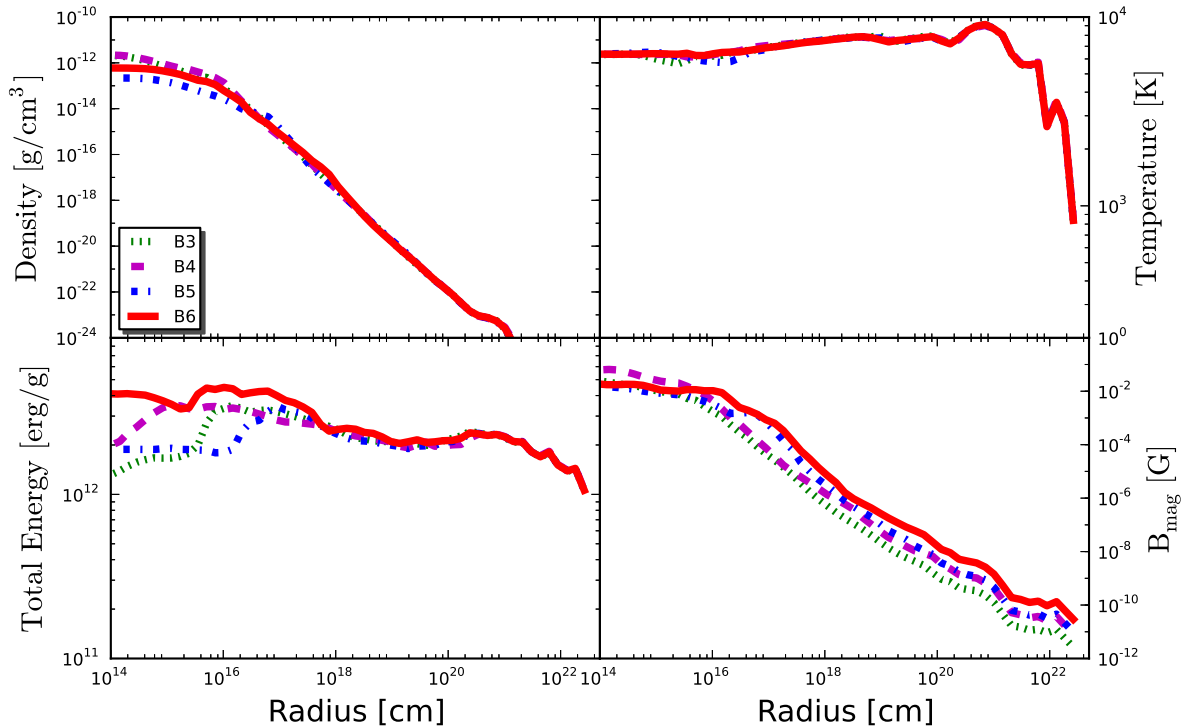}
\caption{ This figure shows the radially binned spherically averaged radial profiles for various initial seed fields as shown in the legend. The upper left panel of the figure shows the density radial profile. The temperature radial profile is depicted the upper right panel. The bottom left panel of the figure shows the total energy radial profile. The radial profile for the total magnetic field strength is shown in the bottom right panel.}
\label{fig5}
\end{figure*}

\begin{figure*}
\centering
\includegraphics[scale=0.5]{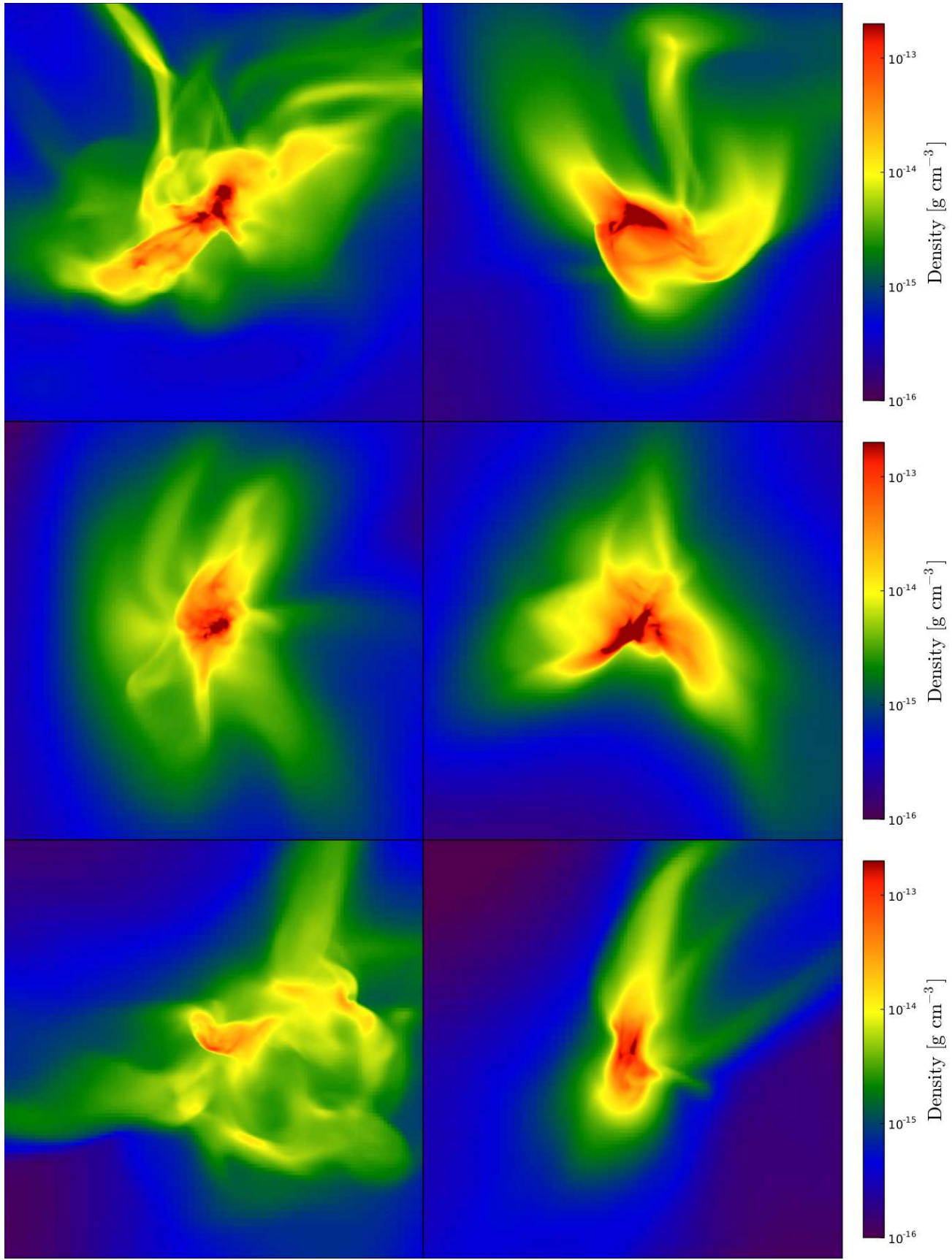}
\caption{Figure shows the density projections at the collapse redshift for various initial seed fields. The central 10000 AU region of the halo is shown for a fixed resolution of 64 cells per Jeans length. The top panels represent B1 (left) and B2 (right), middle panels B3 (left) and B4 (right), and bottom panels B5 (left) and B6 (right).}
\label{fig6}
\end{figure*}

Here, we conducted a number of simulations for different initial fields as listed in table \ref{table1} with a fixed Jeans resolution of 64 cells. The properties of the halo for the different initial seed fields are shown in the figure \ref{fig5}. Similar to the previous cases, the density follows a power law behavior (i.e., $\rm R^{-2}$) and the halo collapses isothermally as seen in the temperature radial profile. The total energy profile is analogous to the earlier cases as depicted in the figure \ref{fig5}. The variations in the interior of the halo arise due to the dissimilarities in the morphology for different field strengths. The strength of the magnetic field for various seed fields is depicted in the bottom right panel. The highest seed field is amplified up to $\rm \sim 0.1$ G, which we suggest to be close to the saturation value (see discussion below). We note that due to minor differences from the Lorentz force, different realizations of the turbulent flow are created, thus changing the morphology as shown in figure \ref{fig6}.

\begin{figure*}
\hspace{-2.0cm}
\centering
\begin{tabular}{c c}
\begin{minipage}{8cm}
\includegraphics[scale=0.5]{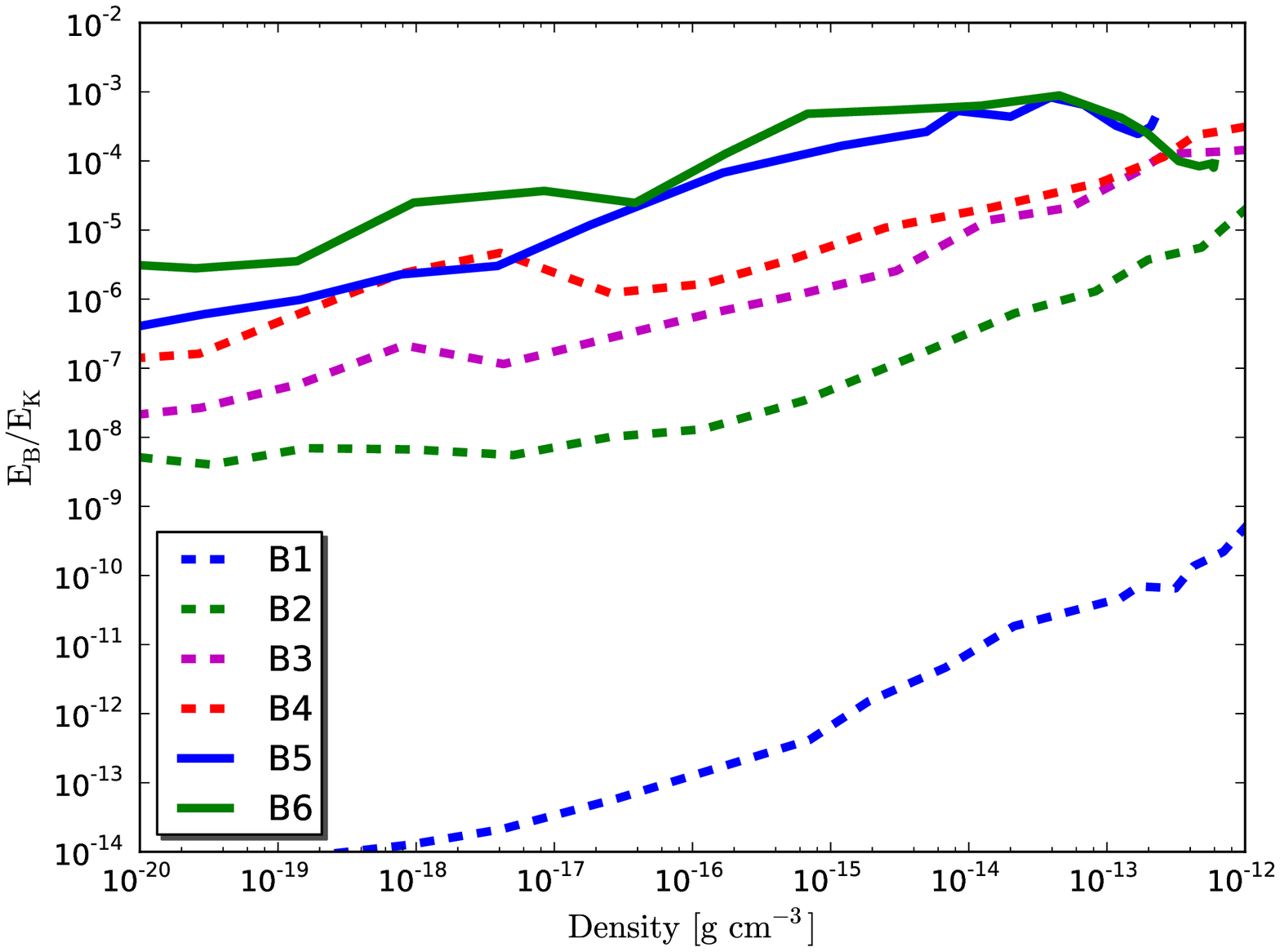}
\end{minipage} &
\hspace{1.0cm}
\begin{minipage}{8cm}
\includegraphics[scale=0.5]{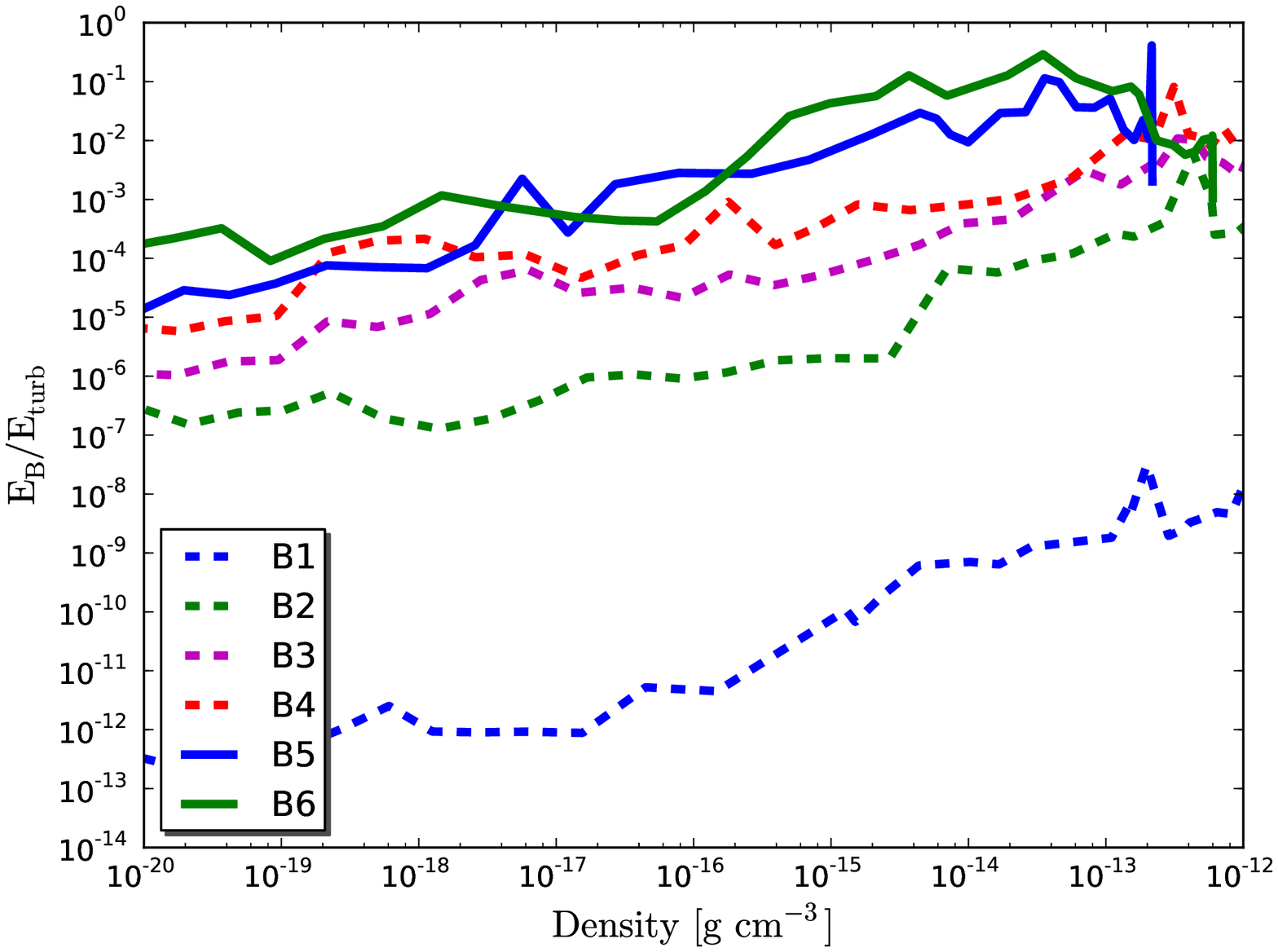}
\end{minipage}
\end{tabular}
\caption{ The left panel shows the ratio of magnetic to kinetic energy plotted. The ratio of magnetic to turbulent energy is shown in the right panel. Different lines represent various initial seed fields as depicted in the legend.}
\label{fig7}
\end{figure*}

To assess the saturation of the dynamo generated magnetic field in a more elaborate way, we have plotted the ratio of magnetic to kinetic energy for these runs which is shown in the left panel of figure \ref{fig7}. This ratio increases with density for all seed fields. We infer the saturation of dynamo generated magnetic field from the change in slope of $\rm E_{B}/E_{K}$, as the latter will change once dynamo amplification stops. The magnetic energy associated with small scale dynamo attains the peak value about $\rm 5 \times 10^{-3}$ the total kinetic energy and then starts to decrease. At that point, we reach approximate equipartition with the turbulent energy. This behavior is observed only for the initial seed field strengths of B5 and B6, while it is absent for the runs B1-B4. In fact for these runs, the ratio of magnetic to kinetic energy keeps rising. We therefore conclude that a transition occurs between B4 and B5, such that the initial field strength is large enough for saturation to occur. To elucidate the saturation of dynamo amplified magnetic field, we have plotted the ratio of magnetic to turbulent energy and is depicted in the right panel of figure \ref{fig7}. The turbulent energy is given by 
\begin{equation}
\rm E_{turb} = {1 \over 2} \rho v_{turb}^{2}~,
\label{eturb}
\end{equation}
where $\rm \rho$ is the gas mass density and $\rm v_{turb}$ is the turbulent velocity. The plot clearly demonstrates that magnetic energy comes close to equipartition with turbulent energy and dynamo amplification gets saturated. Nevertheless, the amplification by gravitational compression will continue until back reactions occur and the system evolves towards a complete saturation. In general, our results are in agreement with previous studies \citep{2011PhRvL.107k4504F, 2012MNRAS.423.3148S}. They also find similar values for the saturation of magnetic energy. 

Our estimates suggest that for a seed field of strength B5 the dynamo generated magnetic field saturates. However, further amplification will continue by gravitational compression. We note that while the dynamo amplification is resolution-limited in numerical simulations, the growth rates in astrophysical gases are considerable higher due to their Reynolds numbers. In the kinematic regime, the growth rate of the magnetic field scales as Re$^{1/3}$ for Burgers turbulence \citep{1999PhRvL..83.2957S, 2005PhR...417....1B}, and Re$^{1/2}$ for Kolmogorov turbulence. In this scenario, the smallest scales will saturate first, and the dynamo growth will then proceed on larger scales. In this non-linear regime, saturation is expected within a few to a few dozen eddy-turnover times \citep{Scheko02, Beresnyak12, Schleicher13}. We thus expect the presence of magnetic fields in approximate equipartition with turbulent energy in the centers of these halos.


\begin{table}
\begin{center}
\caption{The list of simulations performed to probe the saturation of magnetic fields}
\begin{tabular}{cccccc}
\hline
\hline

Model	& Magnetic field strength    \\

 & [$\rm Gauss $]

      \\ 
\hline                                    \\  
 B1	 & $\rm 1 \times 10^{-14}$	\\		
 B2	  & $\rm 6 \times 10^{-12}$	 \\ 
 B3	  & $\rm 1.8 \times 10^{-11}$	   \\ 
 B4	  & $\rm 5.4 \times 10^{-11}$	\\ 
 B5	  & $\rm 6.7 \times 10^{-11}$	 \\ 
 B6	  & $\rm 2 \times 10^{-10}$	   \\
\hline
\end{tabular}
\label{table1}
\end{center}

\end{table}

\section{Summary of main results and Conclusions}

In all, we have conducted 11 cosmological magnetohydrodynamical simulations using the adaptive mesh refinement code Enzo to study the magnetization of massive primordial haloes. We started our simulations with nested grid initial conditions and inserted 27 additional refinement levels during the course of simulations. It gives us an effective resolution down to sub AU scales. To explore the turbulent amplification of magnetic fields in detail, we mandated a Jeans refinement of 16, 32, 64 and 128 cells. These simulations were started at redshift 99 with initial seed field of $\rm 10^{-14}$ G.  We also probed the saturation of dynamo generated magnetic fields for a fixed Jeans resolution of 64 cells. 

In all simulations, the gas falls into the dark matter potential and gets heated up to its virial temperature. At temperatures $\rm \geq 10^{4}$ K, the cooling due to Lyman alpha comes into play and brings the temperature of the gas down to 7000 K. The formation of $\rm H_{2}$ remains suppressed due to the presence of strong photo-dissociation UV radiation flux (i.e., $\rm J_{21}=10^{3}$). Consequently, almost isothermal collapse occurs. We reach a maximum peak density of $\rm 10^{-12}~g/cm^{3}$ in our simulation. The density radial profile follows an isothermal behavior $\rm R^{-2}$ according to the expectation of isothermal collapse. These results are in agreement with previous studies \citep{2002ApJ...569..558O,2009MNRAS.393..858R,2008ApJ...682..745W,2012arXiv1210.1802L} and suggest that thermal properties of the haloes are independent of resolution effects and magnetic seed fields.

We found that the average properties of the halo are almost converged for different resolutions whereas strong dissimilarities in the morphology of the halo are observed for different Jeans resolution. The turbulence is very well resolved for the best resolution run (i.e., $\rm J_{128}$) and will have potential implication for the birth of black holes via direct collapse scenario. Our study suggests higher values of Jeans refinement for atomic cooling haloes (i.e., $\rm \geq J_{64}$) because of higher Mach numbers and larger spatial scales. Previous studies found lower Jeans resolution ($\rm \geq 32$ cells) for minihaloes where turbulence is typically subsonic \citep{2011ApJ...731...62F,2012ApJ...745..154T}. Turbulence causes the medium inside the halo to fragment into clumps and this process is very sensitive to resolution. We stopped simulations because of computational constraints. In future studies techniques like Jeans heating or sink particles will be employed to advance simulation for longer dynamical times scales.

We examined the amplification of magnetic fields and found that it is amplified about ten orders of magnitude. The maximum field strength (i.e., $\rm J_{128}$) is about 100 $\rm \mu$ Gauss. It is initially amplified by gravitational compression under the condition of flux freezing for low resolution runs (i.e, $\rm J_{16}~and~J_{32}$). The role of turbulent amplification at radii larger than 0.1 pc is depreciated due to strong virialization shocks. Turbulence is mainly generated during the gravitational collapse. It drives the small scale dynamo which consequently amplifies the magnetic field strength by randomly stretching, twisting and folding magnetic lines. In the kinematic regime one expects that the growth of magnetic energy is exponential in time while it becomes linear once backreactions of the magnetic field start playing a role. For the highest resolution case (i.e., $\rm J_{128}$), the magnetic field is amplified by at least an order of magnitude above the maximum possible amplification by flux freezing. Indeed, the small scale dynamo gets excited by the gravitational collapse of cosmological initial conditions provided that resolution is enough to capture turbulent 
vortical motions. In general, amplification of the magnetic fields is less efficient compared to the minihaloes because of the higher Mach numbers. Astrophysical systems however have higher Reynolds numbers than those realized in numerical simulations. As the growth rate scales as $\rm Re^{1/3}-Re^{1/2}$, the amplification found here is only a lower limit \citep{2012ApJ...754...99S}. 

We explored the expected saturation values of magnetic fields in MHD simulations by selecting initial seed fields of various strengths. The ratio of magnetic to kinetic energy reaches a peak value of $\rm 5 \times 10^{-3}$ and then starts to decrease when the magnetic energy becomes comparable to the turbulent energy. Further amplification via compression will however occur during gravitational collapse. 

The simulations above confirm that the small-scale dynamo is operating in massive primordial halos, which are generally considered as candidates for the first galaxies and the  formation of massive black holes via direct collapse. Magnetic fields may aid the formation and subsequent growth of massive black holes both via their magnetic pressure and the magnetic tension, providing an additional means for angular momentum transport. While the magnetic pressure is best described via a magnetic Jeans mass, given as $\rm M_{J,B} \propto B^{3}/ \rho^{2}$, the effect of magnetic tension is likely more prominent after the formation of accretion disks. 

The presence of a strong coherent field will further give rise to jets and outflows \citep{Camenzind95, McKinney05, 2012SSRv..169...27P}. These provide a highly efficient mechanism of transporting angular momentum. Considering that the rotation timescale of accretion disks is considerably smaller than the Hubble time, these disks are likely to reach strong coherent fields well before redshift $6$. Such fields may explain the strong rotation measures observed for high-redshift quasars, and potentially enhance the accretion rate to form more massive black holes.

Overall, we summarize our main results as follows:

\begin{itemize}
  \item No turbulent amplification occurs on large scales due to strong virialization shocks.
 \item The turbulent dynamo becomes active in the interior of the halo and amplifies the magnetic field.
 \item The saturation of dynamo generated magnetic field occurs for the initial seed field of $\rm \sim 5 \times 10^{-11}$ G.
\end{itemize}

%
We note that a determination of the saturation level in such cosmological simulations is still uncertain, due to the resolution constraints discussed in the manuscript. It should further be noted that the latter will further depend on the density where saturation occurs. In astrophysical systems, saturation may in particular occur earlier when the density is still lower. In that case, subsequent amplification would likely occur from gravitational compression, thus in the end leading to magnetic field strengths of similar order.

We plan to further explore the consequences of these results in future studies. These will concern in particular the impact of magnetic fields on black hole formation, and potentially its further evolution after the formation of an accretion disk.

\section*{Acknowledgments}
The simulations described in this work were performed using the Enzo code, developed by the Laboratory for Computational Astrophysics at the University of California in San Diego (http://lca.ucsd.edu). We thank the Enzo developers for helpful discussions.  We acknowledge research funding by Deutsche Forschungsgemeinschaft (DFG) under grant SFB $\rm 963/1$, project A12. We thank the HLRN for awarding us computing time via project nip00028. DRGS thanks for funding from the Deutsche Forschungsgemeinschaft (DFG) in the Schwerpunktprogramm SPP 1573 {\em "Physics of the Interstellar Medium"}  under grant SCHL 1964/1-1. The simulation results are analyzed using the visualization toolkit for astrophysical data YT \citep{2011ApJS..192....9T}.

\bibliography{biblio.bib}

\end{document}